# Elucidating the active phases of $CoO_x$ films on Au(111) in the CO Oxidation Reaction

Hao Chen[1], Lorenz J. Falling[2,3], Heath Kersell[2,4], George Yan[5], Xiao Zhao[3,6], Judit Oliver-Meseguer[1], Jaugstetter Max[1], Slavomir Nemsak[2,7], Adrian Hunt[8], Iradwikanari Waluyo[8], Hirohito Ogasawara[9], Alexis Bell[1,10], Philippe Sautet[5,11], and Miquel Salmeron[1,3,6]*

[1] Chemical Sciences Division, Lawrence Berkeley National Laboratory, Berkeley, California 94720, United States;

[2] Advanced Light Source, Lawrence Berkeley National Laboratory, Berkeley, CA 94720, United States;

[3] Materials Sciences Division, Lawrence Berkeley National Laboratory, Berkeley, California 94720, United States;

[4] School of Chemical, Biological, and Environmental Engineering, Oregon State University, Corvallis, Oregon 97331, United States;

[5] Department of Chemical and Biomolecular Engineering, University of California, Los Angeles, Los Angeles, California 90095, United States;

[6] Department of Materials Science and Engineering, University of California, Berkeley, California 94720, United States;

[7] Department of Physics and Astronomy, University of California, Davis, CA 95616, United States;

[8] National Synchrotron Light Source II, Brookhaven National Laboratory, Upton, NY 11973, United States

[9] SLAC National Accelerator Laboratory, 2575 Sand Hill Road, Menlo Park, California 94025, United States;

[10] Department of Chemical and Biomolecular Engineering, University of California, Berkeley, California 94720, United States;

[11] Department of Chemistry and Biochemistry, University of California, Los Angeles, Los Angeles, California 90095, United States.

*Corresponding author. E-mail: mbsalmeron@lbl.gov

# Abstract

Using $CoO_x$ thin films supported on Au(111) single crystal surfaces as model catalysts for the CO oxidation reaction we show that three reaction regimes exist in response to chemical and topographic restructuring of the $CoO_x$ catalyst as a function of reactant gas phase $CO/O_2$ stoichiometry, a finding that highlights the versatility of catalysts and their evolution in response to reaction conditions. Under oxygen-lean conditions and moderate temperatures (≤150°C), partially oxidized films ($CoO_{x<1}$) containing $Co^0$ were found to be efficient catalysts. In contrast, stoichiometric CoO films containing only $Co^{2+}$ form carbonates in the presence of CO that poison the reaction below 300 °C. Under oxygen-rich conditions a more oxidized catalyst phase ($CoO_{x>1}$) forms containing $Co^{3+}$ species that is effective in a wide temperature range. Resonant photoemission spectroscopy (ResPES) revealed the unique role of $Co^{3+}$ sites in catalyzing the CO oxidation. DFT calculations provided deeper insights into the pathway and free energy barriers for the reactions on these oxide phases.

# 1. Introduction

Catalysts are defined as materials that facilitate chemical reactions by providing special sites where reactants and products bind, react, and desorb with low energy barriers separating these steps. Although the composition and structure of the catalyst is usually assumed to be unaltered during reaction many catalysts restructure by displacement of its atoms in response to the adsorption of reactants. [1-3] Here we demonstrate that in addition to topographic restructuring, chemical restructuring can also occur, adding an additional paradigm in the understanding of the working of catalysts. In this work we illustrate this by showing that $CoO_x$ catalysts for the CO oxidation reaction undergo both chemical and topological changes in response to the reactant gas $CO/O_2$ stoichiometry, evolving in three regimes characterized by different Co oxidation states. Under oxygen-lean conditions and moderate temperatures (≤150°C), partially oxidized films of cobalt ($CoO_{x<1}$) deposited on Au containing $Co^0$ were found to be efficient catalysts. With increasing $O_2$ content CoO forms first, which reacts with CO to form carbonates that poison the reaction for temperatures below 300 °C. Finally, under oxygen-rich conditions, more oxidized phases ($CoO_{x>1}$) containing $Co^{3+}$ and $Co^{2+}$ species form that are effective catalysts in a wide temperature range. To follow the catalyst oxidation state, the adsorption of CO, and the reaction products we used Ambient Pressure XPS (APXPS). While the Co oxidation states during reaction can be followed by APXPS, their precise identification and quantification is challenging due to the strong overlap of their 2p core level peaks. We overcame this difficulty using Resonant Photoelectron Spectroscopy (ResPES), which allowed us to precisely identify each Co oxidation state and relative concentration. DFT calculations provided deeper insights into the pathway, stability, and energy barriers for the reactions on each phase.

# 2. Methods

## 2.1 Experimental Methods

APXPS measurements were performed at beamline 9.3.2 of the Advanced Light Source (ALS) at the Lawrence Berkeley National Laboratory, at beamline 23-ID-2 (IOS) of the National Synchrotron Light Source II (NSLS-II) at Brookhaven National Laboratory, and at the Experimental Station 13-2 at the Stanford Synchrotron Radiation Light source (SSRL) at SLAC National Accelerator Laboratory. The Au(111) surface was cleaned by cycles of sputtering (5 min at $3 \times 10^{-5}$ Torr of $Ar^+$ at 1 keV energy), and annealing (10 min at 500 °C) until only Au was detected by XPS. Cobalt films were deposited on the clean Au(111) surface by evaporation from a Co rod (Goodfellow, 99.99+%) using a SPECS e-beam evaporator. The Au substrate was chosen to avoid contributions of the substrate to the reaction, due to its weak binding of CO and inefficient dissociation of $O_2$. The amount of Co was measured by using XPS peak intensities, calibrated with a quartz crystal microbalance (QCM). The amount of Co on the surface is reported in monolayer equivalents (MLE), defined as the amount of Co that would form a complete monolayer if its wetting of Au was perfect. The reported MLE values have an estimated error bar of ±20% (**See SI**). The coverage of $Co^0$ for a given number of MLE was determined using the Co 3p to Au 4f intensity ratio (**Fig. S1**) in comparison with simulated ratios from the quantitative photoelectron simulation package SESSA v2.2[4] (**Fig. S2**). After deposition, the films were oxidized by exposure to $O_2$ at RT or at 200 °C. To ensure high CO purity, the CO gas was passed through a carbonyl trap heated at ~240 °C before entering the measurement chamber. Total gas pressures were monitored with Baratron capacitance pressure gauges. Photon energies of 740, 475, 260, and 920 eV were used to generate photoelectrons with kinetic energies between 150 and 200 eV for the O 1s, C 1s, Co 3p, Co 2p, and Au 4f photoelectrons, respectively, which have mean free paths of ~ 5 Å. Reported binding energies are given with respect to the Fermi level.

*2.2 Computational Methods*

Density functional theory (DFT) calculations were performed using the Vienna Ab initio Simulation Package (VASP) version 5.4.1.[5-7] The exchange correlation energy was calculated using the Perdew-Burke-Ernzerhof (PBE) functional.[8] Spin polarization was used in all calculations. The projector-augmented wave (PAW) method was used to describe the core electrons. The one electron wavefunctions were expanded using a set of plane waves with kinetic energy up to 500 eV. To correct for the self-interaction error of Co 3d electrons, a Hubbard-like onsite repulsion term (DFT+$U$) was included in the calculations using the Dudarev's approach[9], with an effective $U$ value ($U_{eff}$) of 3.5 eV, which has been used in the literature to study the bulk and surface redox properties of $CoO_x$.[10] Structural relaxation for reaction intermediates was performed using the conjugate gradient algorithm. Transition states were first searched using the nudged elastic band (NEB) and climbing image (CI) NEB algorithms.[11, 12] The highest energy image of each CI-NEB calculation was then refined using the Dimer and quasi-Newton algorithms.[13] The electronic structure in each self-consistent field (SCF) cycle was considered converged when the difference in total energy of consecutive steps fell below $10^{-6}$ eV. Atomic positions were considered converged when the Hellman-Feynman forces on unconstrained atoms fall below 0.02 eV/Å. Free energies of CO and $CO_2$ gas were approximated using their translational and rotational partition functions.[14] More details of $CoO_x$/Au(111) structural models are described in the **SI (Fig. S3)**.

## 3. Results and discussion

### *3.1 Deposition, oxidation, and wetting of cobalt films on Au(111).*

Co2p XP spectra from a 1 MLE Co film on Au(111) before and after oxidation are shown in **Fig.1a**. The bottom spectrum (black trace) displays the result before oxidation, showing the $2p_{3/2}$ Co core level peak at 778.2 eV characteristic of metallic Co. After exposing the film to to $10^{-6}$ Torr of $O_2$ gas at room temperature (RT) for 60 seconds (= 60 Langmuir units), a partially oxidized Co film was formed. This is shown in the red spectrum by the additional peak at 780.2 eV, strongly overalpping with the $Co^0$ peak, and its satellite at 786.6 eV, both characteristic of $Co^{2+}$. Fitting the two overlaping $2p_{3/2}$ peaks (**Fig. S4**), we estimate that ~25% of the Co atoms are oxidized to $Co^{2+}$. We will refer to this film as $CoO_{0.25}$. Annealing the film in $1 \times 10^{-6}$ Torr $O_2$ at 200 °C for 10 min led to the formation of CoO, with the peak at 780.2 eV from $Co^{2+}$ now being dominant (**Fig.1a**, blue trace). **Fig.1b** shows the corresponding O 1s XPS region, with the lattice oxygen peak at 529.6 eV, for both $CoO_{0.25}$ and CoO. The peak at 531.2 eV is due to adsorbed OH and CO from residual background $H_2O$ and CO gases.[15] **Fig.1c** shows XPS of the Au 4f and Co3p region after each of these treatments. With an incident photon energy of 260 eV, the kinetic energy of photoelectrons exiting from the Au surface is ~180 eV, with an inelastic mean free path of ~ 5 Å [16] i.e., ~ 2 atomic layers. After deposition of 1 MLE of metallic Co (black trace), the Au 4f peak intensity decreased by ~ 40% compared to the pristine Au(111) (grey trace). This attenuation is consistent with the double-layer island structure of metallic cobalt (see **SI**) as described by Morgenstern et al.[17] After exposing this film to $1\times10^{-6}$ Torr $O_2$ and annealing to 200 °C (blue trace in **Fig.1b,c**), the Au peak intensity decreased to ~ 20% of its clean surface value. This attenuation is consistent with the spreading of $CoO_x$ and can be described well by a layer-by-layer growth (**Fig.S2**).

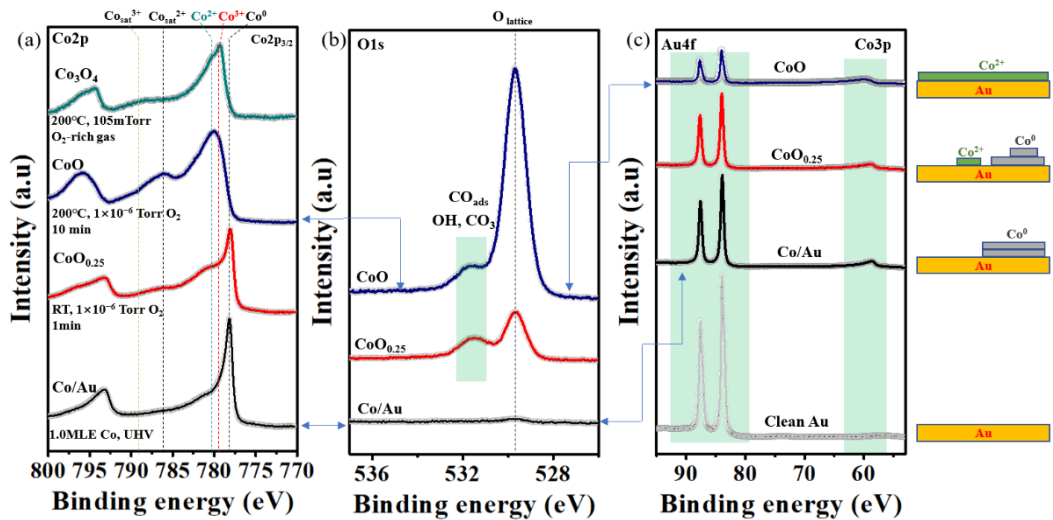

**Fig.1** Oxidation and wetting of cobalt films deposited on Au(111). **(a)** From the bottom: Co2p spectra of 1 MLE $Co^0$ (**black** curve); after room temperature exposure to $1\times10^{-6}$ Torr $O_2$ for 1 min ($CoO_{0.25}$, **red**); after oxidation under $1\times10^{-6}$ Torr $O_2$ at 200 °C for 10 min (CoO, **blue**); and under 105 mTorr of $O_2$-rich gas conditions (CO:$O_2$=1:20) at 200°C ($Co_3O_4$, **green**). The dotted vertical lines mark the binding energy positions of the $2p_{3/2}$ core levels of metallic Co ($Co^0$), CoO ($Co^{2+}$), $Co_3O_4$ ($Co^{2+}$ and $Co^{3+}$), and their shake-up peak satellites ($Co_{sat}^{2+}$ and $Co_{sat}^{3+}$). **(b)** Corresponding XPS O 1s region. From the bottom: mostly clean metallic Co (**black**); $CoO_{0.25}$

film (**red**); and after 10 min exposure to 1×10⁻⁶ Torr $O_2$ at 200 °C (**blue**). **(c)** Au4f core level spectral region. From bottom: clean Au (**grey** curve); after deposition of 1 MLE of Co (**black** curve); after annealing under 1×10⁻⁶ Torr $O_2$ at 200 °C for 10min (**red**); and after 10 min exposure to 1×10⁻⁶ Torr $O_2$ at 200 °C (**blue**). The strong decrease of the Au 4f peak intensity is due to the spreading of the $CoO_x$ film, now covering more Au surface than the initial metallic film. On the right is a schematic illustration of the wetting process during the oxidation process.

Further oxidation at 200 °C in a 105 mTorr of an $O_2$-rich reaction mixture ($O_2$/CO = 20:1) completed the oxidation of cobalt to $Co^{2+}$ and $Co^{3+}$, characterized by peaks at 780.2 and 779.8 eV respectively (green trace in **Fig.1a**). The $Co^{3+}$ is further characterized by the increased intensity of the satellite at 789 eV. We will refer to this film as $Co_3O_4$.[18] The presence of the 2+ and 3+ oxidation states of Co and their proportion in the film will be confirmed and quantified later using ResPES.

### *3.2 CO adsorption on $CoO_{x<1}$ and on $CoO_{x=1}$*

**Fig.2a** shows the C 1s XPS region from the $CoO_{0.25}$ film before and after introduction of 100 mTorr of CO. Before CO introduction (bottom grey trace), the sample shows a C 1s peak at 284 eV, due to adventitious C contamination. In the presence of 100 mTorr of CO, a strong peak at 286 eV appears due to chemisorbed molecular CO (red trace), which adsorbs only on metallic $Co^0$ sites.[19, 20] The small peak at 289 eV is due to carbonate species formed on the CoO areas occupying ~25% of the oxide film area. A peak from gas-phase CO, with its fine vibrational structure is visible at 292 eV. After pumping out the CO gas, the $CO_{ads}$ peak nearly vanished (top gray trace), due to equilibration with the reduced gas pressure. By contrast, on a CoO film under 100 mTorr CO at RT only the peak at 289 eV associated with carbonate species is observed (**Fig.2b**, middle curve), along with the peak of gas-phase CO at 292 eV, but no molecularly adsorbed CO is observed. The area of the carbonate peak on the CoO film became about 4 times larger than that on the $CoO_{0.25}$ film. The amount of carbonates is substantially increased by exposing the sample to $CO_2$ instead of CO, as shown in the top trace in **Fig.2 (b)**.

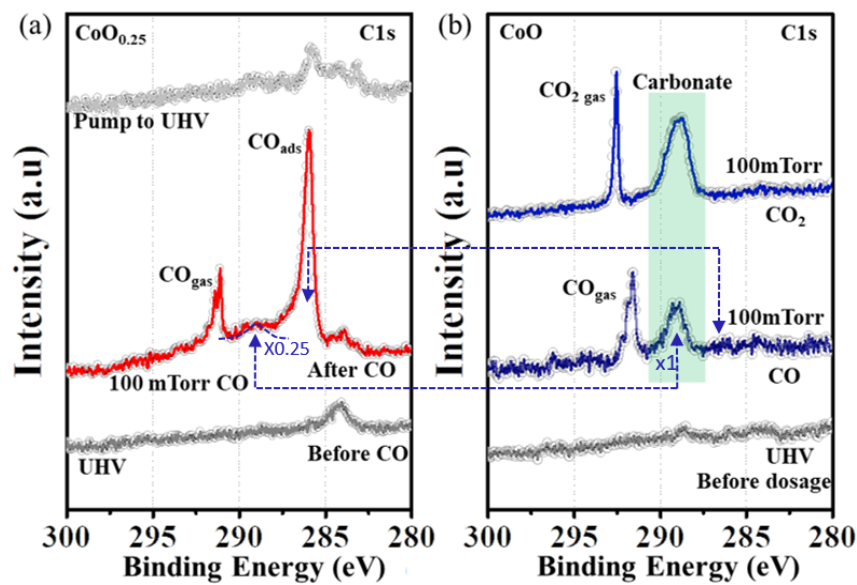

**Fig 2**: Room Temperature CO adsorption on 1MLE of $CoO_{0.25}$ (a) and CoO, on Au(111) (b). (a) C 1s XPS region of $CoO_{0.25}$ in UHV (**gray**, bottom); under 100 mTorr of CO (**red**, middle), and after pumping out the gas (**gray**, top); **(b)** C1s XPS region of CoO in UHV (**gray**, bottom), under 100 mTorr CO (**blue**, middle) and 100 mTorr $CO_2$ (**blue**, top). Molecular CO adsorbs only on metallic Co, and forms carbonates on the CoO regions. A copy of the peak shape of the carbonate peak in (b) marked x1, is superimposed at the same BE position in the red curve in (a) after reducing its height by a factor of 4 for comparison. The amount of carbonate increases substantially in the presence of $CO_2$ gas (top trace).

### *3.3 CO oxidation reaction catalyzed by partially oxidized cobalt ($CoO_{x<1}$)*

In the previous section we identified the species formed by CO adsorption on $CoO_{x<1}$ and $CoO_{x=1}$ at RT. Here we follow the evolution of the $CoO_{0.25}$ surface during the CO oxidation via the Mars van-Krevelen mechanism as a function of temperature. The surface composition, followed by APXPS, is shown **Fig.3**. In UHV, and in 100 mTorr of CO at RT the spectra in the C 1s region are similar to those in **Fig. 2.** Heating to 100 °C caused a decrease in the intensity of the C 1s peak at 286 eV from adsorbed CO due to the new equilibrium with the gas phase at the higher temperature. This is shown by the O lattice peak at 529.6 eV and the $Co^0$ peak at 778 eV (**Fig. 3(b,c))**, which remained essentially unchanged at this temperature. The O 1s peak at 531.8 eV, due to overlaping O peaks from adsorbed CO and OH, decreased due to thermal desorption. Raising the temperature to 150 °C increased the reaction rate, as shown by the increase of the $Co^0$ peak at 778.2 eV, the decrease of Co $2p_{3/2}$ peak at 780.2 eV from $Co^{2+}$ (**Fig.3a),** and the decrease of the lattice oxygen peak at 529.6 eV. All these changes confirm the reduction of CoO$_{0.25}$ by reaction between CO and lattice oxygen.

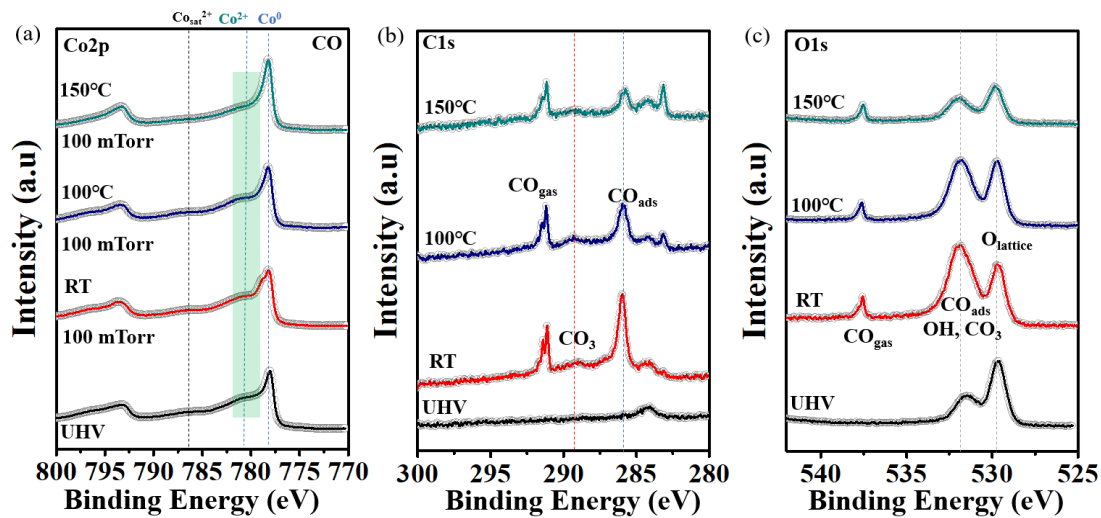

**Fig.3.** APXPS in the (a) Co 2p (b) C 1s, and (b) O 1s core level regions acquired during the CO oxidation reaction on Au-supported 1 MLE CoO$_{0.25}$ film. From bottom: at RT in UHV (black), under 100 mTorr of CO at RT (red), at 100 °C (blue), and at 150 °C (green). Above 100 °C the reduction of Co$^{2+}$ in the CoO$_{0.25}$ film is shown by the increase in the Co 2p$_{3/2}$ peak and the decrease of the O$_{latt}$ peak near 530 eV. The reaction rate is still low due to the presence of carbonate (peak at ~289 eV) that blocks the reaction.

### *3.4 CO reactions on cobalt monoxide (CoO)*

After oxidation of 1 MLE of Co to form CoO, with all cobalt atoms in the Co$^{2+}$ state, CO was introduced in the chamber to a pressure of 100 mTorr. The reaction was monitored by APXPS as a function of temperature, with the results shown in **Fig.4**. The spectra in panel (a) show that the intensity of the Co$^{2+}$ and O peaks remained largely unchanged up to 300°C but dropped rapidly thereafter. The decrease of these peaks can be attributed to the combined oxide reduction upon decomposition of the carbonate and CoO dewetting and that creates CoO clusters that exposes more Au surface, as shown by the rapid increase of the Au 4f peak intensity in **Fig. 4(d).** This is and illustrated schematically in the inset. The deweting is the result of the formation of unstable carbonates, and reveals a topographical restructuring accompanying the change in oxidation state phases, from CoO$_{x=1}$ to CoO$_{x<1}$ . We discuss this further in the theoretical **Section 3.6**.

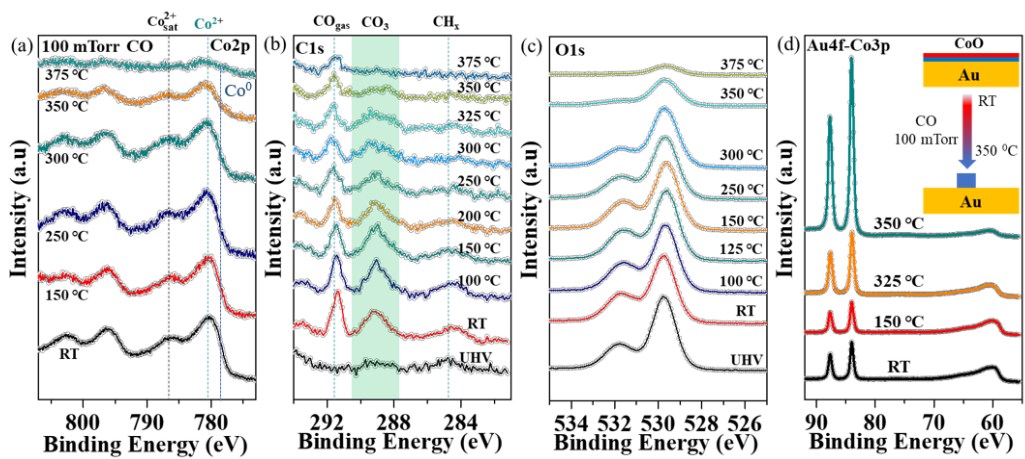

**Fig.4.** Reduction and de-wetting of CoO/Au by reaction with CO: (a-c) Co 2p, C 1s, O 1s, and Au 4f XP spectra of 1 MLE of CoO on Au. From bottom: in UHV (black), and under 100mtorr CO after heating to the temperatures indicated. The decrease in the intensity of the Co and O peaks above 300 °C is related to reduction, clustering, and dewetting of the unstable carbonate covered CoO, as shown by the rapid increase of the Au 4f peak shown in (d). Inset: graphic illustration of the dewetting process.

### 3.5 CO oxidation catalyzed by $CoO_{x>1}$

In the previous sections, the oxide film contained $Co^{\delta+}$ species with $\delta \leq 2$. However, under oxygen-rich conditions, oxide phases including O-Co-O trilayers [18, 21] and $Co_3O_4$ islands[22], containing $Co^{3+}$ are present, which has been proposed by several authors to be the active site in oxygen-rich conditions.[23-26] To our knowledge however, this has not been proven spectroscopically in operando conditions. To ascertain this important point, it is necessary to unambiguously distinguish spectroscopically the Co oxidation states, $Co^{3+}$ and $Co^{2+}$, both involving different partially filled and empty d-levels. This can be done by Resonant Photoelectron Spectroscopy (ResPES) as proposed and demonstrated by several groups.[27, 28] Briefly, ResPES is based on the photoemission of electrons from d-band states, enhanced by the resonant excitation of core level electrons to empty d-states that decay by an Auger process of energy equal to the initial X-ray. This is illustrated on the left panel of **Fig.5** for one of the oxidation states.[29, 30] As the electronic configuration differs between $Co^{3+}$ ($[Ar]3d^6$) and $Co^{2+}$ ($[Ar]3d^7$), different resonant excitation energies for 2p to 3d transitions exist for each species. Since Auger-mediated emission and direct photoemission are undistinguishable process with identical final states, a strong resonant enhancement of the photoemission spectra is observed. The photon energy for the resonant excitation can be experimentally determined by collecting valence band photoemission spectra as a function of X-ray energy. A maximum emission will be obtained at the resonant energy.

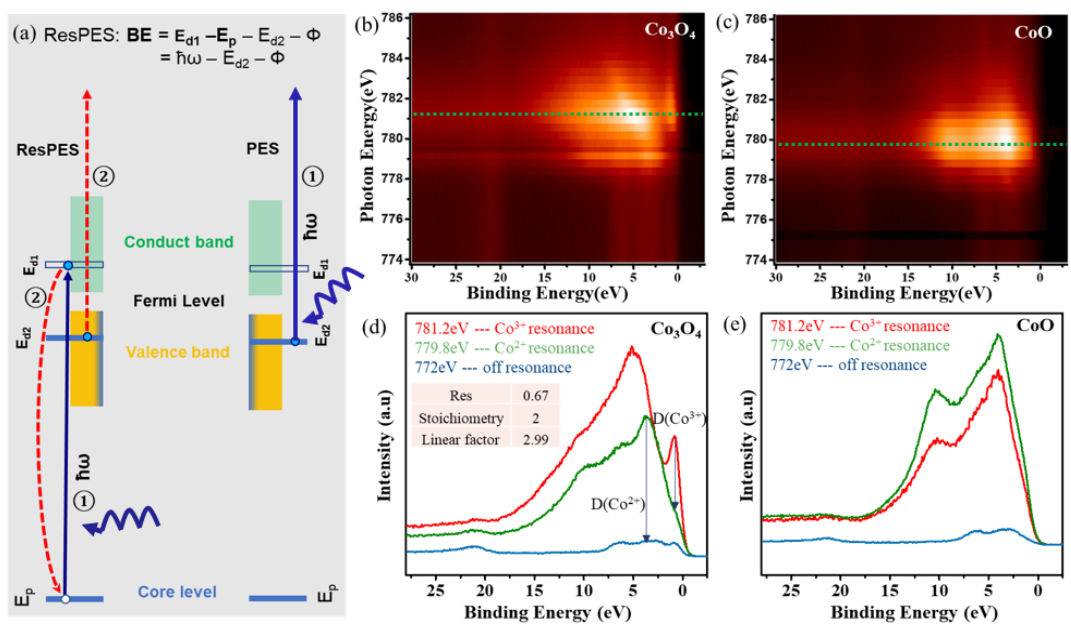

**Fig.5** (a) Illustration of the ResPES process: a photon excites an electron from a core level $E_p$ to an unoccupied d1 level (**blue** arrow, left) belonging to a particular Co oxidation state, which decays by an Auger process (**red** lines) ejecting an electron from an occupied d2 state of that ion (left). The same final state can be obtained by direct absorption of the resonant photon by electrons in the same d2 level (right). These two processes are undistinguishable and thus interfere to enhance the transition. **(b-c)** Photoemission heat maps of the Valence Band (VB) states from 1 MLE of $Co_3O_4$ and of 1MLE of CoO on Au (X=Binding Energy, Y=Exciting photon energy, Z= color coded photoemission intensity). **(d-e)** Resonant VB spectra of $Co_3O_4$ and CoO along the dashed green lines in (b) and (c) for the resonant energies of 781.2 eV, and 779.8 eV. The bottom VB spectra (blue) were acquired with off-resonance photons of 772 eV energy. The valence d-levels of the Au substrate contribute also to the spectra.

The experimental determination of the resonant energies for 1 MLE of $Co_3O_4$, containing $Co^{3+}$ and $Co^{2+}$ on Au is shown in the heat maps (bright to dark for high and low intensity) of **Fig.5b-c**. The valence band spectra at the resonant energies of 781.2 eV for $Co^{3+}$ and 779.8 eV for $Co^{2+}$ (dashed green lines in (b) and (c)) are shown in **Fig. 5d-e**, with the green curves corresponding to the resonant spectra for $Co_3O_4$ and CoO. Peaks for the filled d-levels of the two oxidation states are visible around 10 eV, 5.0 eV and 1.0 eV for Co3O4 and around 10.5 eV and 4.0 eV for CoO. As can be seen they are strongly enhanced compared with the off-resonance density of states acquired using 772 eV photons (blue spectra at the bottom), which contains also the 5d peaks of the valence band of the Au substrate at approx. 2.5, 4 and 6 eV.[31]

The contribution from the $Co^{3+}$ and $Co^{2+}$ ions, $D(Co^{3+})$ and $D(Co^{2+})$ in **Fig.5(d)**, can be quantified by the difference in peak intensities relative to the $Co^{2+}$ and to the off-resonance spectra respectively, and can be used to determine the relative concentration of these species because the resonant enhancement ratio (RER), $D(Co^{3+})/D(Co^{2+})$, is directly proportional to the ratio of the concentrations [27], $N(Co^{3+})/N(Co^{2+})$, with a linear correction factor (y) that can be determined from the known ratio in stoichiometric Co3O4 films, through the equation:

$$\frac{N(Co^{3+})}{N(Co^{2+})} = y * \frac{D(Co^{3+})}{D(Co^{2+})} = y * RER$$

From **Fig.5d-e** the RER of stoichiometric Co3O4 is 0.67. Since the $N(Co^{3+})/N(Co^{2+})$ value for stoichimetric Co3O4 (Co2O3-CoO) is 2, the linear factor y = 2.99. Therefore, we can determine the concentration ratio, $N(Co^{3+})/N(Co^{2+})$ of nonstoichiometric $CoO_x$ films through the measurement of the RER.

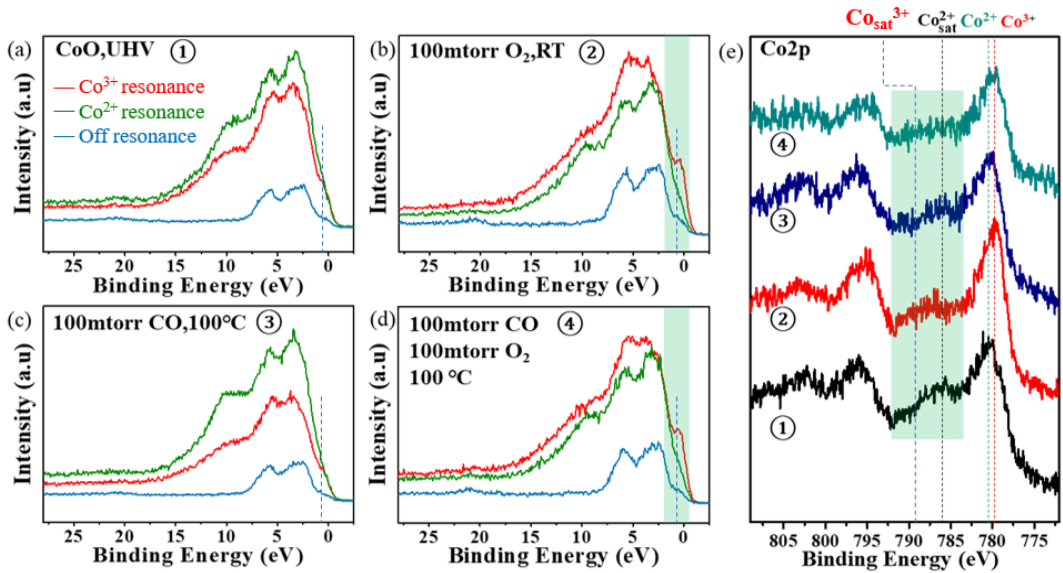

**Fig.6** Resonance Photoemission Spectra (ResPES) collected during the CO oxidation reaction on a 0.5 MLE $CoO_x$ on Au(111) sample. **(a-d)** ResPES for an X-ray of energy ℏω =781.2 eV ($Co^{3+}$ resonance, **red** curve), for ℏω =779.8 eV ($Co^{2+}$ resonance, **green** curve) and for 772 eV (off-resonance, **blue** curve). **(e)** Co 2p XP spectra of CoO in UHV, black curve; under 100 mTorr O2 at RT (**red** curve); under 100 mTorr CO at 100°C (**blue** curve); and under 100 mTorr CO and 100 mTorr O2 (**green** curve). In (a) the dashed line marks the $Co^{3+}$ peak position in the valence band and in the XPS. The intensity of the resonant peak increases from (a) to (b) due to oxidation, and decreases (c) due to reduction by reaction with CO. The peak is present when both CO and $O_2$ are present during reaction as shown in (d). **(e)** APXPS for each of the reaction conditions in (a to d)).

The ResPES spectra in **Fig.6a-d** demonstrate the role of $Co^{3+}$ species in 0.5 MLE of $CoO_{x>2}$ on Au(111) in the CO catalytic oxidation reaction. The ResPES in **Fig.6a** corresponds to the initial film in UHV where the film is not completely oxidized (Fig.**6e**, black curve). Exposure to 100 mTorr $O_2$ at RT, increased the peak of $Co^{3+}$ at 778.6 eV in the XPS (**Fig.6e**, red curve), more clearly evidenced by the rise of the characteristic resonant peaks at 10 eV, 5.0 eV and 1.0 eV in **Fig.6b**. The RER of this intermediate $CoO_x$ is 0.3 and, therefore its $N(Co^{3+})/N(Co^{2+})$ ratio is nearly unity. The oxide intermediates remained structurally stable at RT when the reaction conditions were switched to 100 mTorr CO (**Fig. S5**). The nonstoichiometric cobalt oxide was reduced to CoO when the reaction temperature was raised to ~100 °C, which is clearly seen in the ResPES of **Fig.6c**. The reduction of $CoO_x$ at 100 °C indicates the high reactivity of the intermediate oxide phase. Importantly, adding 100 mTorr of $O_2$ to the 100 mTorr of CO at 100°C regenerated the intermediate phase as indicated by the re-appearance of the $Co^{3+}$ resonant peak at ~1.0 eV (**Fig.6d**) and the red shift of the Co $2p_{3/2}$ peak (**Fig.6e**, green curve). Since this nonstoichiometric oxide phase is structurally stable under CO and $O_2$ mixtures (CO/$O_2$ = 1), it appears to be the most catalytically active phase for CO oxidation at mild tempertures in oxygen-rich conditions. From its RER value of 0.47, the $N(Co^{3+})/N(Co^{2+})$ ratio of this active phase is 1.4, indicating a higher ratio of $Co^{3+}$ species at 100 °C compared with that at RT. Furthermore, the stoichiometry parameter "x" in this intermediate $CoO_x$ is 1.29, namely $CoO_{1.29} \approx Co_3O_{3.87}$, which is structurally equal to the $Co_3O_4$ phase. $Co^{3+}$ species were also proposed to be the active sites in $Co_3O_4$ nanorods for the cryogenic CO oxidation reaction.[25] Recent surface science results have also confirmed the facile activation of molecular $O_2$ to peroxide ($O_2^{2-}$) and superoxide species ($O_2^-$) at the oxygen vacancies on the surface of $Co_3O_4$(100) model catalysts[26]. This suggests that the newly-formed $Co^{3+}$ in the reactive environment, together with oxygen vacancies, is responsible for the enhanced catalytic reactivity of our Au(111)-supported nonstoichiometric $CoO_x$ monolayer films.

### *3.6 DFT calculations of the reaction between CO and $CoO_x$/Au*

To further understand the reactivity of $CoO_x$ structures on Au(111), DFT calculations of the reaction between CO and surface O were performed on $CoO_{x<1}$/Au(111), CoO/Au(111), and $CoO_{x>1}$/Au(111) structural models (**Fig. S3**).

For $CoO_{x<1}$/Au, like on $CoO_{x<1}$/Pt[15], CO adsorbs exergonically by -0.35 eV atop a Co atom not bound to O, in agreement with the presence of molecular CO detected by APXPS (**Fig. S6**). The adsorbed CO can react with two types of surface O located at either a Face Centered Cubic site (FCC) or a Hexagonal Close Packed site (HCP) sites (with respect to surface Co). The reaction between adsorbed CO and surface O is endothermic, by 1.35 eV for O at at the FCC site and 1.17 eV for O at the HCP site, which agrees with the required heating of the $CoO_{x<1}$/Au film before reaction between lattice CO and surface O could take place. Next, the formation of carbonate groups on CoO/Au was investigated by DFT calculations. It was found that to be unlikely for carbonate groups to form at the terraces of a stoichiometric bilayer CoO films, because the formation of $CO_2$ was calculated to be just as exothermic as that of $CO_3^{2-}$ (**Fig. S7**). We tested the sensitivity of carbonate group formation to changes in Co-Co spacing by performing the same calculations over a CoO film with a narrower 3.00 Å Co-Co spacing, but the same trend was observed. Since the structure of bulk $CoCO_3$ is known [32], we examined the possibility that the formation of carbonates induced the restructuring of the oxide layer. The calculations indicate that a flat $CoCO_3$ overlayer on Au(111) is unstable and weakly bound. The optimized structures were found to dewet and relax away from the Au surface (**Fig. S8**). This is supported by the results in Fig. 4d where the deweting of the oxide upon heating above 150°C and beyond is revealed by the rapid increase in the Au4f XPS peak. Thus, it is likely that the experimentally observed carbonate groups are located on a restructured CoO terrace [18, 33] or that they form at the edge of CoO islands which retain their 2D structure, as previously reported for CoO films on Pt(111).[15]

Finally, to understand the superior reactivity of CO oxidation on $CoO_{x>1}$/Au,

DFT calculations were performed to obtain the free energy barriers of the CO-lattice O reaction (**Fig.7a** and **Fig. S9**). Two possible structures for the $CoO_{x>1}$/Au were evaluated: a $Co_3O_4$ film **(Fig. 7)** and a $CoO_2$ film **(Fig. S9)**. Following our previous calculations of the structure of $Co_3O_4$(111) surfaces, a Co-poor and O-rich surface was chosen to simulate a $Co_3O_4$(111) film.[34, 35] On this termination, only Co cations originally in bulk tetrahedral sites and O anions are exposed. To initiate the reaction, CO binds weakly on exposed Co. The adsorbed CO can react readily with O atoms bound to Co neighbors by crossing a 0.49 eV barrier. At 100 °C and 100 mTorr of CO, the net free energy barrier for CO-lattice -O reaction is 1.04 eV. We note that this barrier is lower than the CO-lattice O reaction energy over $CoO_{0.5}$ (1.17 eV, **Fig. S5b**), making $Co_3O_4$ more reactive. On the other hand, the $CoO_2$/Au(111) film appears less reactive than $Co_3O_4$ as a 1.24 eV initial barrier is required for the reaction between CO gas and lattice O (**Fig. S9**). Weakly-bound $CO_2$ produced by this exothermic reaction can either desorb or further react with lattice O by crossing a 0.04 eV barrier **(Fig.7b)**. The $CO_3^{2-}$ group formed in this step is a bidentate species interacting with adjacent exposed Co. Although easy to form, the $CO_3^{2-}$ group can also decompose by crossing a 0.71 eV barrier (**Fig.7b**). Carbonates formed by reaction between $CoO_2$ and the $CO_2$ product are also easy to decompose, requiring an even smaller barrier of 0.21 eV (**Fig. S9**). Under a low $CO_2$ partial pressure ($P_{CO2}$ = 0.1 mTorr), the exergonic $CO_2$ desorption prevents the easily formed $CO_3^{2-}$ from poisoning the surface.

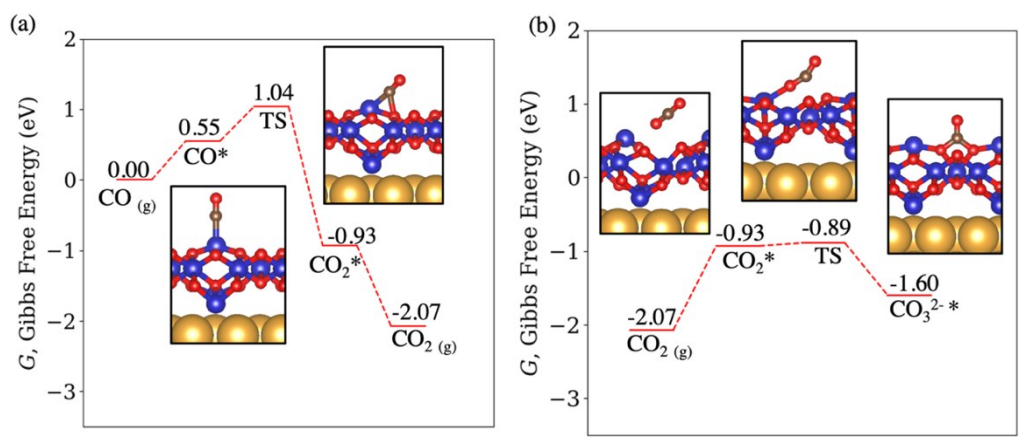

**Fig.7** Free energy pathways of the reaction of (a) CO and (b) $CO_2$ with $Co_3O_4$/Au(111). CO gas reacts readily with $Co_3O_4$ by crossing a 1.04 eV free energy barrier relative to CO gas. The formation of carbonates by reaction between lattice O and $CO_2$ gas is unfavorable as carbonates will readily decompose (right-to-left in panel b). The free energies of gas CO and $CO_2$ were calculated at 373.15 K, $P_{CO}$ = 100 mTorr, and $P_{CO2}$ = 0.1 mTorr. Color scheme: Au: yellow; C: brown; Co: blue; O: red.

## Conclusions

The structure and reactions of Au-supported ultrathin $CoO_x$ catalyst films in the presence of $CO/O_2$ mixtures, was studied with the goal of determining the evolution of the catalyst structure and the active sites involved in CO oxidation. It was found that $CoO_{x<1}$ films have a high chemical reactivity towards CO oxidation under oxygen-lean conditions at mild temperatures (from 100 °C to 150 °C), which we attribute to the presence of O vacancies near $Co^{\delta+}$ where CO can adsorb and react with neighboring lattice oxygen. In contrast, on stoichiometric CoO, CO adsorbs forming carbonates that poison the CO oxidation. The formation of carbonates on the surface and their persistence up to ~300 °C, where their decompositon rate is rapid, divides the reaction into two regimes of high catalytic activity. One is characteristic of oxygen-lean conditions containing both $Co^0$ and $Co^{2+}$, the other is characteristic of oxygen-rich conditions containing $Co^{3+}$ species that are reduced by CO to $Co^{2+}$ and regenerated by $O_2$ back to $Co^{3+}$ at temperatures of 100 °C and below. Using resonant X-ray photoemission spectroscopy (ResPES) we demonstrated unambigously the role of $Co^{3+}$ as an active site and the catalyst phase under reaction conditions. DFT calculations indicate that the high reactivity is due to a lower energy barrier for C-O bond formation. Our findings provide a general understanding of the enhanced catalytic reactivity of cobalt oxide catalysts. They underline paradigm of a double restructuring of the catalyst, one is chemical (i.e. Co oxidation state), the other topographical (CoO detachment and dewetting from the Au substrate), which is induced by reactant composition and temperature.

## Acknowledgments

This work was supported by the Office of Basic Energy Sciences (BES), Chemical Sciences, Geosciences, and Biosciences Division, of the U.S. Department of Energy (DOE) under Contract DE-AC02-05CH11231, FWP CH030201 (Catalysis Research Program). L. J. F. acknowledges support from the Alexander von Humboldt Foundation, Bonn, Germany. It used resources of the Advanced Light Source, a U.S. DOE Office of Science User Facility under contract no. DE-AC02-05CH11231, the 23-ID-2 (IOS) beamline of the National Synchrotron Light Source II, a User Facility operated for the DOE Office of Science by Brookhaven National Laboratory under Contract No. DE-SC0012704, and the Stanford Synchrotron Radiation Light Source, SLAC National Accelerator Laboratory, supported by the U.S. Department of Energy, Office of Science, Office of Basic Energy Sciences under Contract No. DE-AC02-76SF00515. X.Z. was supported by NSF-BSF grant number 1906014. The DFT calculations in this work used computational and storage services associated with the Hoffman2 cluster at the UCLA Institute for Digital Research and Education (IDRE), and the Bridges-2 cluster at the Pittsburgh Supercomputing Center (supported by National Science Foundation award number ACI-1928147) through the Extreme Science and Engineering Discovery Environment (supported by National Science Foundation grant number ACI-1548562) grant TG-CHE170060.